\documentstyle[prl,aps,twocolumn]{revtex}

\begin{document}

\wideabs{
\title{Individual addressing and state readout of trapped ions utilizing rf- micromotion}

\author{D. Leibfried}

\address{Inst. f. Experimentalphysik, Universit\"at Innsbruck, Technikerstr. 25, A-6020 Innsbruck, Austria}
\date{\today}
\maketitle

\begin{abstract}
A new scheme for the individual addressing of ions in a trap is described that does not rely on light beams tightly focused onto only one ion. The scheme utilizes ion micromotion that may be induced in a linear trap by dc offset potentials. Thus coupling an individual ion to the globally
applied light fields corresponds to a mere switching of voltages on a suitable set of compensation electrodes. The proposed scheme is especially suitable for miniaturized rf (Paul) traps with typical dimensions of about 20-40 microns.
\end{abstract}

}

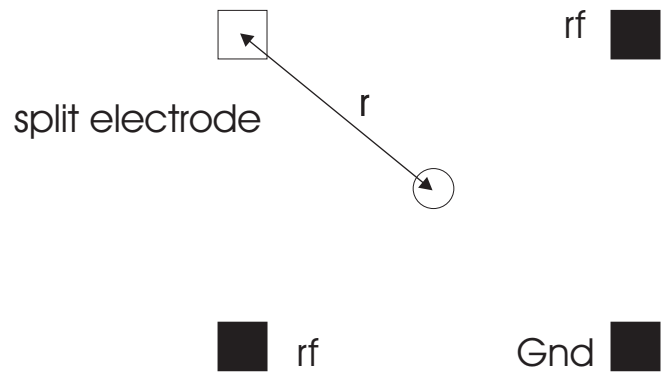
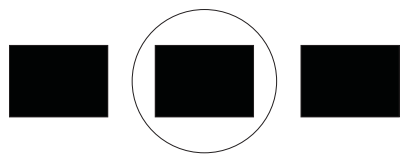
\begin{figure}

\caption{\label{schema} (a)Schematic drawing of the two ground rods of the trap. (b) Cut through the trap. The rf-node line is at distance $r$ from the split ground rod. (c) Approximation of the electrostatic situation used in the model calculation. The rod sections with spacing $d$ are replaced by spheres of radius $d/2$.} 
\end{figure}

Even the realization of elementary quantum information processing operations puts severe demands on the experimental techniques. A single two qubit quantum gate, for instance, requires two strongly interacting quantum systems, highly isolated from environmental disturbances. In 1995 Cirac and Zoller proposed a realization of quantum logic gates using a string of ions trapped in a linear rf (Paul) trap \cite{Cirac95}. In the meantime several experimental steps towards quantum logic gates implemented with trapped ions have been demonstrated, namely cooling one \cite{Monroe95a} and two ions \cite{King98} to the ground state, a quantum CNOT-gate connecting the internal and motional states of one ion \cite{Monroe95b} and the deterministic creation of entangled states \cite{Turchette98}.\\
In the Cirac-Zoller proposal addressing individual ions is done by tight focussing of a laser beam on one and only one ion. In this case inter-ion distances must be bigger than the diffraction limit of the addressing laser beam and cannot become smaller than roughly one micron. This requirement limits the minimum size of the trap and also the maximum level spacing of the harmonic trapping potential, since the inter-ion distance is determined by the balance of mutual Coulomb repulsion and the strength of the external potential \cite{Wineland98}. The trap level spacing in turn limits the maximum speed of gate operations because all laser pulses must be long enough to discriminate between different motional levels. The limitation in inter-ion distances also rules out motional frequencies beyond the linewidth of allowed dipole transitions commonly used for Doppler-cooling in ion traps. Such motional frequencies would allow for very efficient cooling without the need for special techniques like Raman-cooling\cite{Monroe95a}.\\
An alternative addressing technique that does not rely on focussed laser beams could therefore be very advantageous because it lifts restrictions in trap size, cooling and processing speed inherent in the Cirac-Zoller scheme. In \cite{Turchette98}, state creation was based on the dependence of transition-frequencies of individual ions on the micromotion induced by the trapping ac-field. This letter describes how this basic idea might be extended to individually address and read out single ions in a trap without the need for tightly focused laser beams.\\
After a good compensation of micromotion in a linear trap \cite{Berkeland98} the whole string of ions will reside on the rf-node line and will therefore not undergo forced oscillations in the rf-trapping field. With a suitable static electric field one and only one ion may be pushed from the rf node line, while leaving the others there. Since this one ion now undergoes forced oscillations at the rf drive frequency $\Omega_{\rm rf}$, its two internal levels may be coupled on one of the {\it micromotion} sidebands with a Rabi-frequency $\Omega_{\rm m}$ proportional to the first Bessel function $J_{1}({\bf k} \cdot {\bf{\xi}})$\cite{Turchette98} where ${\bf \xi}$ is the amplitude vector of the driven motion and ${\bf k}$ is the wavevector (or wavevector difference for Raman-transitions) of the driving light field. The other ions reside at the node line and are therefore not coupled, although the coupling laser beam might illuminate them. Similarly the final state of individual ions may be read out by detecting fluorescence induced on a micromotion sideband of a dipole allowed cycling transition. The method turns out to be especially suitable for micro-fabricated traps with typical dimensions of 20 to 40 $\mu$m. A linear quadrupole trap of this type with a distance between diagonally opposite electrodes of 600 $\mu$m was successfully demonstrated by the NIST group \cite{Myatt98} and no technical limits seem to forbid further miniaturization by a factor of 10-20.\\
One way to create enough degrees of freedom for the 'individual compensation' sketched above is to split one of the ground rods of the linear quadrupole trap into a number $N_{s}$ of sections that is greater or equal to the number $N_{i}$ of ions trapped (see Fig. \ref{schema} a). The sections are individually wired to control their voltages independently. As specified below in the case $N_{s}=N_{i}=N$ and equally spaced sections of length $d$ one can balance the individual compensation voltages in such a way, that only one of the ions is pushed into a position where the rf-field is nonzero, while all the other ions remain on the rf field-node line.\\    

To further elaborate how this might be done, it will be assumed that all ions are initially on the rf-node line. In the experiment this means the micromotion of all ions has to be carefully nulled with the same set of electrodes \cite{Berkeland98} and then the displacement voltages introduced below will be added to the null voltages. The electric field created by an individual compensation electrode will of course depend on its actual shape. For the purpose of this letter it will be sufficient to approximate compensation electrode $j$, held at voltage $U_{j}$ by the field created by a conducting sphere of radius $d/2$ held at that voltage (see Figs. \ref{schema} b, c). The magnitude of the electric field $E_{j}(r)$ at distance $r$ is then
\begin{equation}
E_{j}(r)= \frac{U_{j} d}{2 r^2}.
\label{field}
\end{equation}
Inter-ion spacing can be controlled to some extend by the dc voltages applied to the trap endcaps \cite{James98}. Here it will be assumed that the ions are equally spaced with an inter ion distance $d$ and are a distance $r$ away from the electrodes (See Fig \ref{schema} a, b). This approximation is valid for 3 ions and for the center part of a long string. One could use the exact inter ion distances \cite{James98}, but this would only complicate the following discussion without radically changing the underlying physics. The ions and electrodes are labeled from left to right. The electric field of electrode $j$ at the position of ion $i$ can then be decomposed into a part parallel to the trap axis 
\begin{equation}
E^{(z)}_{ij}=\frac{U_{j} d^2}{2 (r^2+[(i-j)d]^2)^{3/2}},
\label{epar}
\end{equation}
and the perpendicular part that will actually push the ion into the rf-field
\begin{equation}
E_{ij}=\frac{U_{j} d r}{2 (r^2+[(i-j)d]^2)^{3/2}} = m_{ij} \frac{U_{j} d}{2r^2},
\label{eperp}
\end{equation}
with relative distance factors $ m_{ij}=(1+[(i-j)d/r]^2)^{-3/2}$.
The total field perpendicular to the trap axis experienced by ion $i$ is therefore
\begin{equation}
E_{i}=\sum_{j=1}^{N} E^{t}_{ij}.
\end{equation}
With respect to the slow (secular) motion, the field will just change the equilibrium positions of the ions but will have no effect on the vibrational modes (in the limit of small amplitudes) \cite{James98}. The perpendicular part will displace an ion of mass $m$ and charge $Q$ off the trap axis by an amount \cite{Berkeland98}
\begin{equation}
y_{i}=\frac{8 Q E_{i}}{m q^2 \Omega_{\rm rf}^2},
\label{displ}
\end{equation} 
where $q=(2 Q V)/(m r^2 \Omega_{\rm rf}^2)$ and $V$ is the amplitude of the rf-field. At this position the ion will undergo micromotion with an amplitude of \cite{Berkeland98}
\begin{equation}
\xi_{i}=y_{i}\frac{q}{2} = \frac{2 E_i r^2}{V}.
\end{equation}
The index $\kappa_{i}={\bf k} \cdot {\bf \hat{y}} \xi_{i}$ (${\bf \hat{y}}$ is a unit vector along the y-direction) with which an incoming laser beam will now be frequency modulated in the rest frame of ion $i$ depends on its wave vector $\bf{k}$ (or the wave vector difference ${\bf \delta k}={\bf k_{1}}-{\bf k_{2}}$ for Raman excitation) and will therefore be determined by the actual experimental geometry. In the remaining ${\bf k}$ is assumed to be parallel to ${\bf \hat{y}}$,
\begin{equation}
\kappa_{i}={k \xi_{i}}= k y_{i}\frac{q}{2}.
\label{modind}
\end{equation}
Reexpressing Eq.(\ref{modind}) in terms of the compensation voltages in  Eq.(\ref{eperp}) yields:
\begin{equation}
\kappa_{i} = \frac{k d}{V}\sum_{j=1}^{N} m_{ij} U_{j}.
\end{equation}
This defines a system of linear equations for the compensation voltages $U_{i}/V$ expressed in multiples of the rf-amplitude $V$ that can be written in matrix form:
\begin{equation}
{\bf m} \cdot \frac{{\bf U}}{V} = \frac{\kappa}{k d} {\bf e},
\label{matrix}
\end{equation}
with $\kappa_{i}=\kappa e_{i}$. The condition that only ion $l$ is modulated with index $\kappa$ while all other ions are unmodulated corresponds to a solution of Eq. \ref{matrix} with ${e_{i}=\delta_{il}}$. Small corrections to the actual ion positions due to the compensation field component along $z$ [Eq. (\ref{epar})] and the slight change in the mutual repulsion of the ions could be incorporated in a self consistent way, but are neglected in the first order approximation presented here. Solutions of Eq. (\ref{matrix}) may be found analytically for small ion numbers or numerically. 

As one might expect, the scaled amplitudes $U_{j}/V$ become very large if $r \gg d$, because then the difference in magnitude of the matrix elements $m_{ij}$ is very small. The method is therefore more suited for traps where $r$ is within an  order of magnitude with the ion spacing $d$. In the examples calculated below $r=$15~$\mu$m and an ion spacing $d$ of 3~$\mu$m is used. These parameters could be realized by scaling the existing micro-fabricated trap of the NIST group \cite{Myatt98} down by a factor of 20,  and lowering the rf-amplitude $V$ by a factor of 400 compared to the values used in that trap. The Rabi frequency on one of the first micromotion sidebands will be $\Omega_{m} = J_{1}(\kappa) \Omega_{0}$ where $\Omega_{0}$ is the unmodulated carrier Rabi frequency. 

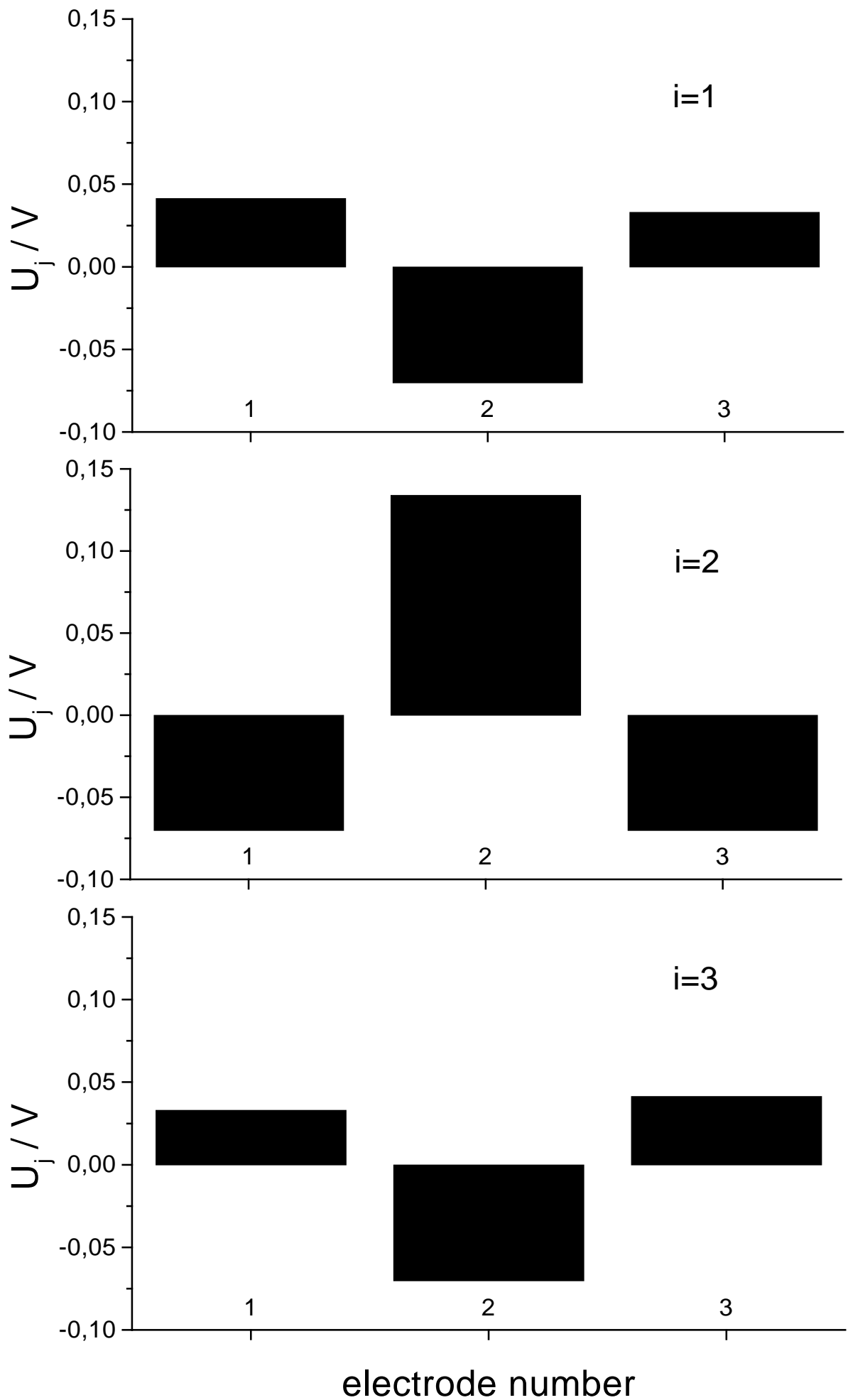
\begin{figure}

\caption{\label{ions3} Scaled compensation voltages $U_{j}/V$ versus electrode number $j$ for (top to bottom) $i$=1,2,3.}
\end{figure}

Results calculated with $J_{1}(\kappa)=$0.1 which corresponds to $\kappa \simeq$0.2 and $k= 2 \pi/(313{\rm nm})$ resonant with the $^9$Be$^+$ S-P transition are shown in Fig. \ref{ions3} for 3 ions and Fig. \ref{ions51} for 10 and 51 ions. Figure \ref{ions3} shows the scaled voltages $U_{j}/V$ ($i$=1,2,3) for the aforementioned parameters, while in figure \ref{ions51} a) $U_{j}/V$ for addressing ion 5 of 10 ions is displayed. Fig. \ref{ions51} b) is a plot of the magnitude of the electric field perpendicular to the trap axis generated by the 10 electrodes held at the voltages given in Fig. \ref{ions51} a). As expected there are 9 positions where the axial field is zero, coinciding with the ion positions of all but ion 5. At this position the field exhibits a maximum. Finally Fig. \ref{ions51} c) shows the electrode voltages necessary to address the middle ion of a string of 51 ions. The voltages seem to be easy to realize in the case of three ions (peak voltage 0.15 V for an rf-amplitude of 2.5 V, roughly 1/400 of the voltages used in the Boulder micro-trap \cite{Myatt98}) and manageable in the case of 51 ions (peak voltage 30 V for the same rf amplitude). Ion numbers between 4 and 50 will require voltages between 0.25 and 30~V. The maximum ion displacements are on the order of 40~nm along the trap axis and the addressed ion $i$ is displaced roughly 200 nm perpendicular to the trap axis. These displacements could be further diminished by operating the trap at a higher $q$ parameter (e.g. $q=$0.6).    

In conclusion a new way to individually address single ions in a linear trap that is suitable for quantum information processing was proposed. A simple calculation yielded the order of magnitude of all relevant parameters. All required conditions could be met with existing technologies.
The proposed mechanism of addressing a quantum gate reduces a severe optics and engineering problem to a  mere changing of voltages on compensation electrodes. No diffraction limited focussing on single ions and complicated schemes to switch the beam on and off or shuttle it around in the trap are necessary. In principle the coupling laser could illuminate the ions all the time during a computation, while the 'pulse lenght' of the interaction is solely controlled by switching the compensation voltages. The scheme renders the distance limit for inter-ion spacing set by the laser wavelenght in the Cirac-Zoller scheme obsolete and the modifications to push e.g. a special pair or more ions off the rf-node line and thus realize multi-ion gates are obvious. The read-out of individual ions can be done the same way as the addressing, since the micromotion sideband frequency may be chosen much higher than the natural linewidth of typical cycling transitions used in ion-traps. Motional heating was found to be substantially suppressed in the non center-of-mass vibrational modes \cite{King98}. In those modes Rabi-frequencies of individual ions depend on their position. This dependencies could readily be incorporated into the definition of $\kappa_i$ so operations on all ions would require the same interaction time and laser intensity. Finally reaching the Lamb-Dicke regime with Doppler cooling is straightforward in miniaturized traps so one could use the gate operations proposed by S{\o}rensen and M{\o}lmer \cite{Sorensen98} which also work with the ions in thermal motion and in the presence of moderate motional heating. This implies that the experimentally most demanding preconditions of the Cirac-Zoller proposal, namely ground state cooling and addressing by focussed laser beams, may be circumvented in future quantum logic experiments.\\

The author is grateful for discussions with D. J. Wineland, J. C. Bergquist, C. Monroe, C. Myatt, Q. Turchette, H. C. N\"agerl, C. Roos and R. Blatt. He also acknowledges a EC TMR postdoctoral fellowship under contract ERB-FMRX-CT96-0087.

\begin{figure}

\caption{\label{ions51} a) Scaled compensation voltages $U_{j}/V$ versus electrode number $j$ for 10 ions and $i$=5, b) Perpendicular component of the electric field for the compensation voltages of a). Note that all ions but the fifth reside at a node of the field. c) Scaled compensation voltages for 51 ions and $i$=26.}
\end{figure}

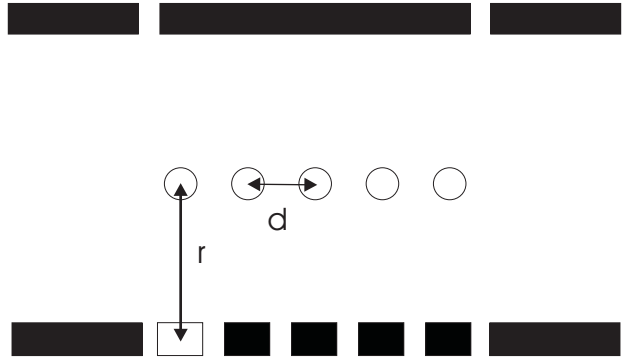
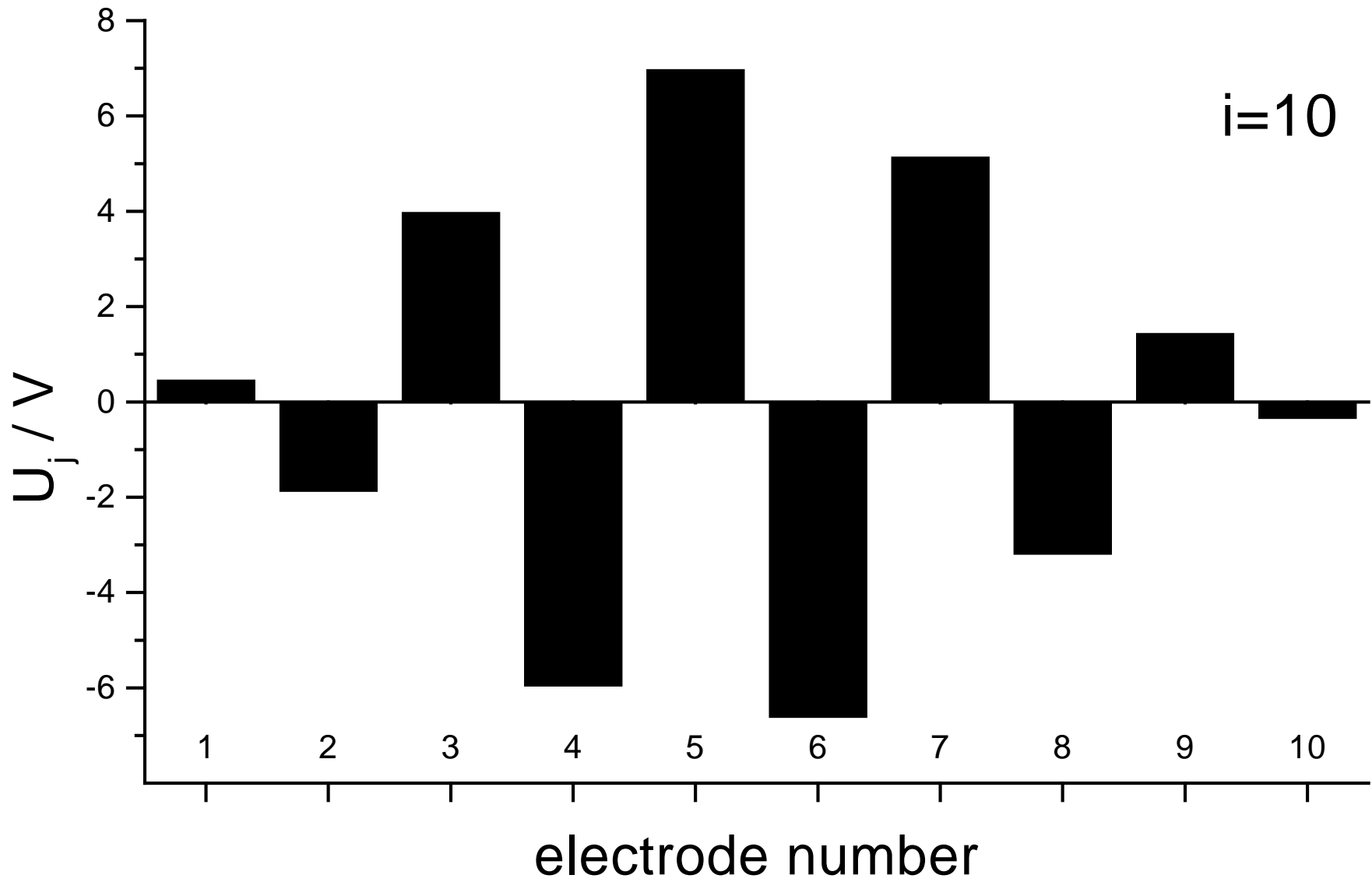
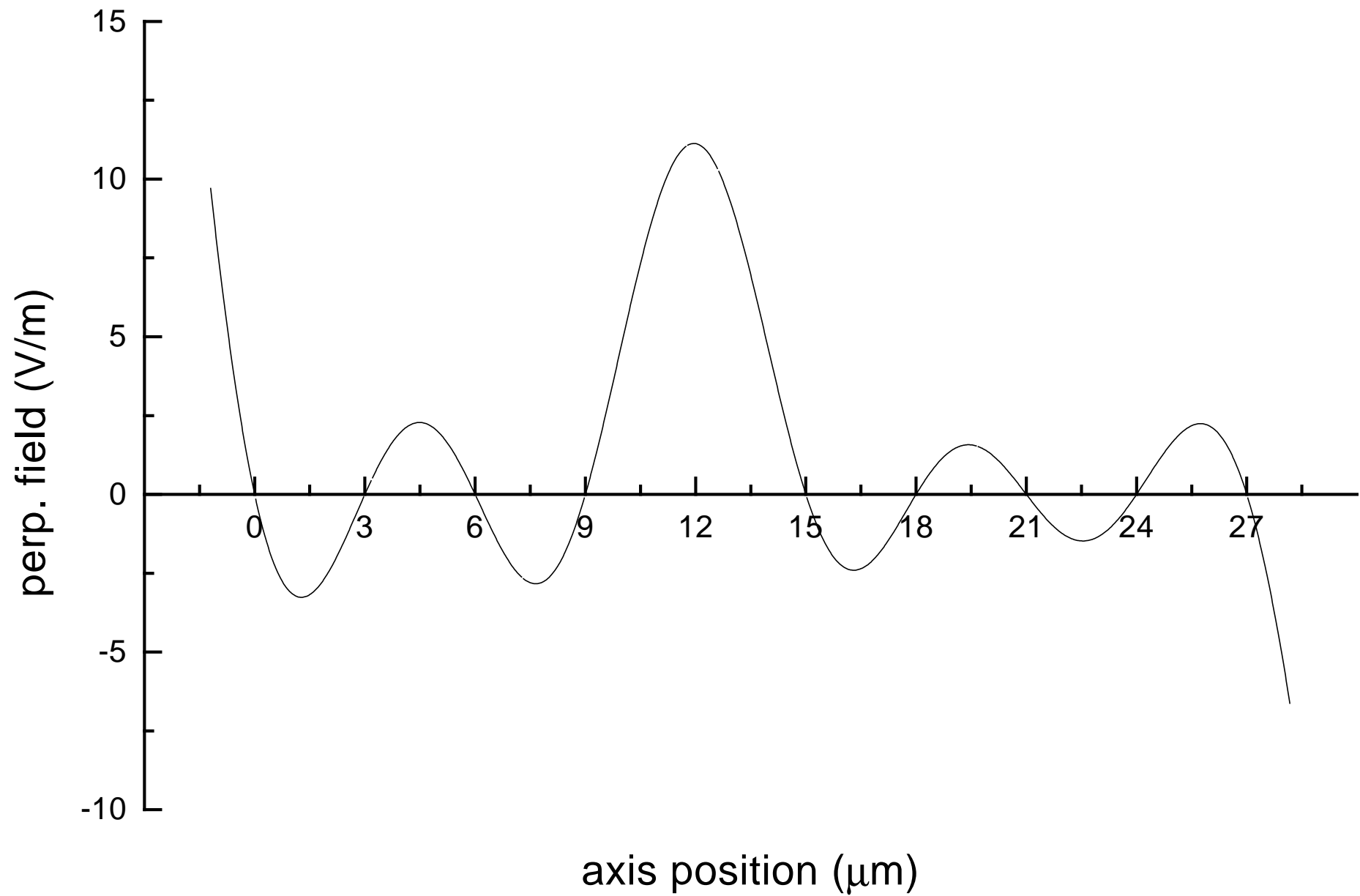
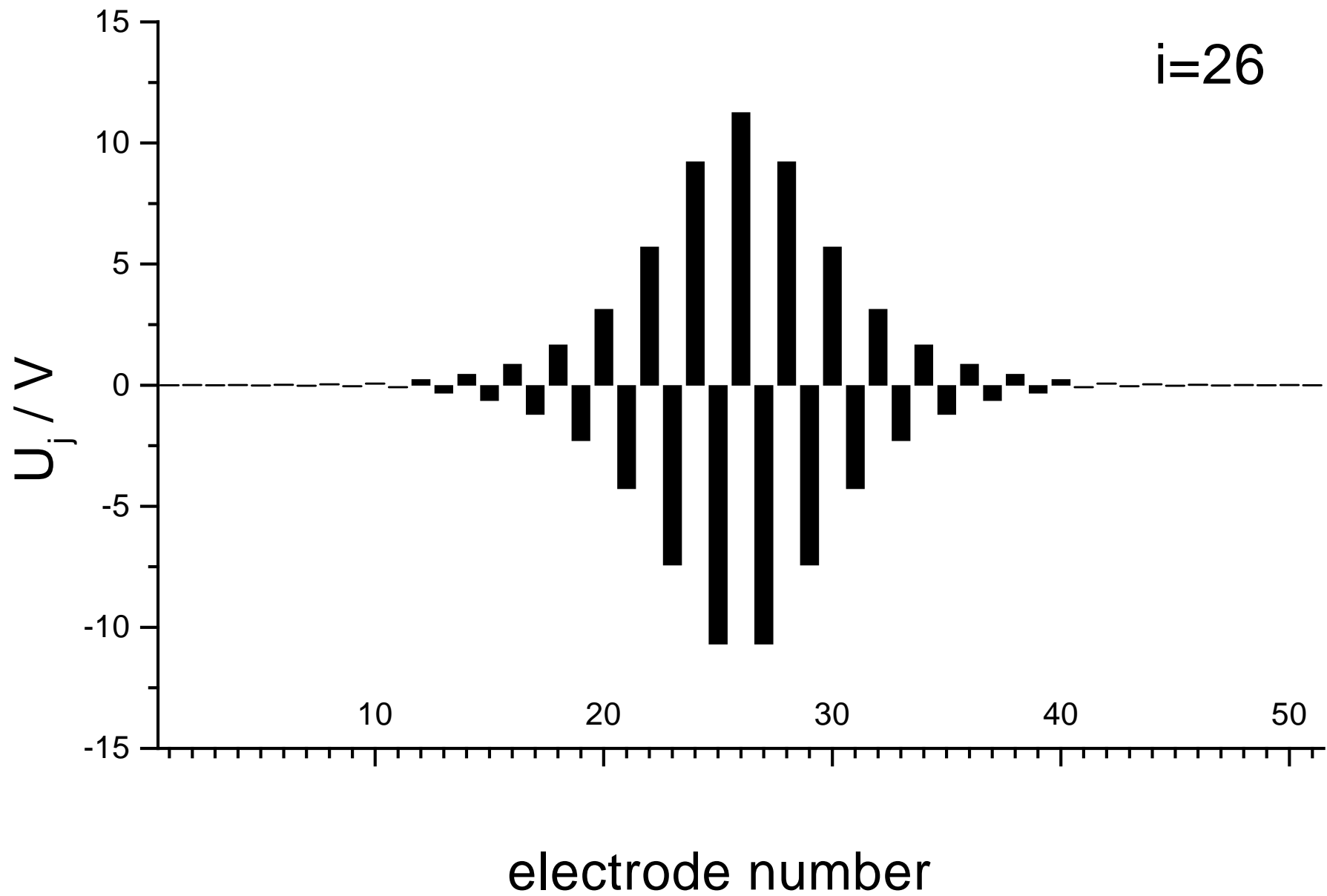
\end{document}